\documentclass[review]{elsarticle}

\usepackage{lineno}
\usepackage{url}

\usepackage{breakurl}
\usepackage[breaklinks]{hyperref}
\modulolinenumbers[5]

\journal{Journal of \LaTeX\ Templates}
\usepackage{epsfig,amssymb,latexsym,color,amsmath,pifont,colordvi,multicol}
\usepackage{subfigure}
\usepackage{amssymb}
\usepackage{dsfont}
\usepackage{setspace}
\usepackage{bbm}
\usepackage{wrapfig}
\usepackage{algorithm}
\usepackage{algpseudocode}
\usepackage[thinlines]{easytable}
\usepackage{stackengine}

\definecolor{backgrey}{rgb}{0.86,0.86,0.86}
\definecolor{dblue}{rgb}{0,0.0,0.5}
\definecolor{dred}{rgb}{0.4,0.2,0}
\definecolor{dgreen}{rgb}{0.0,0.5,0}

\newcommand{\captionfonts}{\small}
\makeatletter  
\long\def\@makecaption#1#2{%
  \vskip\abovecaptionskip
  \sbox\@tempboxa{{\captionfonts #1: #2}}%
  \ifdim \wd\@tempboxa >\hsize
    {\captionfonts #1: #2\par}
  \else
    \hbox to\hsize{\hfil\box\@tempboxa\hfil}%
  \fi
  \vskip\belowcaptionskip}
\makeatother   
\newtheorem{theorem}{Theorem}

\newtheorem{assumption}[theorem]{Assumption}

\newtheorem{definition}[theorem]{Definition}

\newcommand{\A}{\bf \alpha}
\newcommand{\X}{\bf X}
\newcommand{\Y}{\bf Y}

\begin{document}

\begin{frontmatter}

\title{Causal Analysis and Prediction of Human Mobility in the U.S. during the COVID-19 Pandemic}

\author{Subhrajit Sinha*}\cortext[mycorrespondingauthor]{Corresponding author}
\ead{subhrajit.sinha@pnnl.gov}
\author{Meghna Chakraborty}
\ead{chakra43@msu.edu}
%

\fntext[fn1]{Pacific Northwest National Laboratory. 
902 Battelle Blvd, Richland, Washington, 99354}
\fntext[fn2]{Michigan State University. 428 S Shaw Ln. East Lansing, Michigan, 48910}

\begin{abstract}
Since the increasing outspread of COVID-19 in the U.S., with the highest number of confirmed cases and deaths in the world as of September 2020, most states in the country have enforced travel restrictions resulting in sharp reductions in mobility.  However, the overall impact and long-term implications of this crisis to travel and mobility remain uncertain.  To this end, this study develops an analytical framework that determines and analyzes the most dominant factors impacting human mobility and travel in the U.S. during this pandemic.  In particular, the study uses Granger causality to determine the important predictors influencing daily vehicle miles traveled and utilize linear regularization algorithms, including Ridge and LASSO techniques, to model and predict mobility.  State-level time-series data were obtained from various open-access sources for the period starting from March 1, 2020 through June 13, 2020 and the entire data set was divided into two parts for training and testing purposes.  The variables selected by Granger causality were used to train the three different reduced order models by ordinary least square regression, Ridge regression, and LASSO regression algorithms.  Finally, the prediction accuracy of the developed models was examined on the test data.  The results indicate that the factors including the number of new COVID cases, social distancing index, population staying at home, percent of out of county trips, trips to different destinations, socioeconomic status, percent of people working from home, and statewide closure, among others, were the most important factors influencing daily VMT.  Also, among all the modeling techniques, Ridge regression provides the most superior performance with the least error, while LASSO regression also performed better than the ordinary least square model.
\end{abstract}


\end{frontmatter}


\section{Introduction and Background}\label{section_intro}

The novel coronavirus (COVID-19) pandemic is delineating our times' global health crisis and has had a significant impact on the way we understand our everyday lives.  Since its emergence in Asia in late 2019, all continents in the world except Antarctica have been fighting the virus in earnest.  The World Health Organization (WHO) \cite{WHO} has declared COVID-19 a global pandemic on March 11, 2020 and the United States declared a national emergency on March 13, 2020 \cite{whitehouse}.  As of September 15, 2020, almost 30 million cases of COVID-19 were confirmed in 215 countries around the globe \cite{worldometer} and among all countries, the U.S. has the highest number of confirmed cases and fatalities in the world due to COVID-19 \cite{worldometer}.  Several countries have closed their borders, exercised lockdowns, curfews, stay-at-home orders, and social distancing protocols, resulting in sharp reductions in mobility and travel demand at local, regional, national, and even international levels.  By March 24, 2020, more than 20 percent of the world’s population has been ordered to remain at home as governments, health, and administrative organizations take extreme measures to protect their communities from the spread of the virus \cite{guardian}.

In the U.S., more than 40 states have already enforced stay-at-home order, the earliest being in California effective from March 19, 2020 \cite{stay-at-home_CA}.  Only in the U.S., the mobility restrictions during the COVID-19 pandemic, in the form of travel bans, stay-at-home mandates, and lockdown policies, have impacted millions of people.  Overall, human mobility has been severely impacted due to the travel restrictions and individual concerns to avoid public gatherings, resulting in tremendous economic impacts in transportation sectors.  However, the overall influence and the long-term implications of this pandemic to mobility and transportation systems still remain unknown at this point in time.  Against the background of this unprecedented global crisis, questions remain as to how the different factors during the pandemic affect human mobility and travel.

With the increasing availability of high-quality data related to COVID-19, analyzing transportation and mobility during and after this crisis is imperative.  Although there are still many unknowns, statistical models and analytical tools would help produce evidence-based research and policy interventions after COVID-19 outbreak.  At the national level, Zhang et al. (2020) developed a COVID-19 impact analysis platform \cite{UMDCOVID:2020} that can inform users about the spread of COVID-19 in the U.S., and the effects of the virus spread and government orders on mobility and social distancing in the country, using privacy-protected smartphone device location data, coupled with the census information.  The platform gets updated daily and goes back to January 1, 2020, for benchmarking, and the results are scaled and aggregated to the entire population for both state and county levels \cite{zhang2020interactive}.  Gao et al. (2020) reported on the interactive web-based mapping platform \cite{UWMadisonPlatform} developed by the GeoDS Lab at the University of Wisconsin, Madison, with the support of the National Science Foundation RAPID program \cite{NSF_link}.  This platform provides information on how people in different counties and states in the U.S. responded to the social distancing directives and guidelines.  The platform integrates geographic information systems (GIS) and the daily updated human mobility patterns obtained from large-scale anonymous and aggregated smartphone data at the county-level \cite{gao2020mapping}.

In recent times, research in several domains is heavily focusing on large-scale data analysis utilizing sophisticated computing capabilities and machine learning \cite{golbeck2011predicting,zheng2015big,nguyen2018deep, chakraborty2020linear,  chakraborty2020safety, chakraborty2021examining, erickson2017machine}.  However, causal analysis of data for prediction purposes has received limited attention in the literature.  Causality has been one of the oldest and under-examined questions in all of science. However, conceptually, its importance has been widely acknowledged in prior studies and used in various scientific disciplines including social media research \cite{IT_social_media}, neuroscience \cite{IT_brain}, biological networks \cite{IT_bionetwork1,IT_bionetwork2}, and economics and finance \cite{granger1969,sims_money,granger1988}, among others.  Causal analysis and influence characterization were initially geared towards time-series analysis and among the different measures, Granger causality \cite{granger1969,granger1988}, directed information \cite{IT_massey_directed,IT_kramer_directedit}, and Schreiber's transfer entropy \cite{IT_schreiber} have gained the most attention.  In the realm of dynamical systems, a new definition and measure of causal characterization, called information transfer, has been proposed recently \cite{sinha_IT_cdc_2015,sinha_IT_cdc_2016,sinha_IT_icc}, where the authors show that the existing definitions of causality, namely, Granger causality, directed information, and transfer entropy fail to capture the correct causal structure in a dynamical system.  Additionally, some recent studies \cite{sinha_IT_data_acc,sinha_IT_data_jnls} provided a data-driven approach to infer the causal structure of a dynamical system from time-series data.

Although there are different measures of causality, for time-series data analysis, Granger causality is one of the simplest and most widely used methods.  Broadly, in econometric studies, Granger causality test has been popular in analyzing time-series data for identifying influential factors and prediction purposes.  Konstantakis et al. (2017) employed Granger causality, along with other quantitative techniques, to investigate the factors that affect automobile sales in Greece \cite{konstantakis2017modeling}.  More recently, Homolka et al. (2020) determine macro and socio-economic indicators that may significantly predict car registrations across European countries with the help of Granger causality test and Vector Autoregressive models \cite{homolka2020short}.  Also, in the domain of transportation research, particularly in predictive analysis with big data, Granger causality has been employed widely.  Beyzatlar et al. (2014) analyzed panel data from fifteen European countries (EU-15) using Granger causality to investigate the relationship between income and transportation  \cite{beyzatlar2014granger}.  In order to accurately build the traffic flow prediction models, Li et al. (2015) utilized Granger causality to determine the potential dependence among the pool of predictor variables in the time-series big data collected by different sensors \cite{li2015robust}.  In a relevant study, McMullen and Eckstein (2012) tested Granger causality between vehicle miles traveled (VMT) and various measures of the national economic activity over time \cite{mcmullen2012relationship}.  The authors rightly argued that it is imperative to properly understand the relationship between VMT and other pertinent transportation and economic factors, as VMT trend is one of the key components in transportation policies, in general.

Moreover, regression analysis is commonly used in data analysis and machine learning \cite{mitchell_book}, and is applied extensively in various domains, including transportation \cite{chakraborty2020linear, chakraborty2021association, chakraborty2021analysis, chakraborty2021assessing, chakraborty2021relationship}.  Among the various regression algorithms, linear regression is computationally one of the most efficient techniques and is often used as a starting point for many different problems, although in many cases, the linear regression model may be suboptimal.  In the analysis of time-series data coming from a dynamical system, linear regression arises naturally in data-driven analysis of dynamical systems using the transfer operators, namely Perron-Frobenius and Koopman operators \cite{rowley2009spectral,williams2015data,sinha_sparse_koopman_acc,sinha_koopman_equivariant}.  More recently, studies \cite{robust_koopman_acc,robust_koopman_journal} used robust optimization techniques to compute the aforementioned operators from noisy data sets, and the resulting optimization problem was a variation of the ordinary least squares (OLS), namely, least squares with regularization.  Regularization is a standard technique used in data analysis to overcome some of the limitations of OLS, including the overfitting and susceptibility to noise in the data \cite{mitchell_book}.

The literature on transportation analysis utilizing regularization is scant.  Recently Polson and Sokolov (2017) developed a deep learning model utilizing a linear model that is fitted with $\ell_1$-regularization to predict traffic flows.  The study showed that deep learning architecture was capable of capturing non-linear spatio-temporal effects in traffic and providing short-term predictions of traffic flow \cite{polson2017deep}.  Tan et al. (2011) proposed a semi-supervised Elastic Net regression method for pedestrian counting by utilizing sequential information between unlabelled samples and their temporally neighboring samples as a regularization term.  The developed model was able to attain superior prediction performance and select representative features from the original set of features without losing their interpretability \cite{tan2011semi}.  Hasan et al.  (2017) proposed statistical techniques to identify spatial relationships among road links in an urban road network to select predictors for a short-term traffic prediction model for a given road link.   The study used a time-lagged multiple linear regression method and utilized two analytical methods, including the Granger causality test and Elastic Net regularization, using one year of traffic flow and speed data from the selected road network in Brisbane, Australia.  For a given target link, the relevant predictors obtained by the Granger causality and Elastic Net were used separately to build the respective traffic prediction models.  The results showed that Granger causality-based traffic prediction model provided superior prediction accuracy than that using the Elastic Net regression \cite{hasan2017spatial}.  More recently, Battifarano and Qian (2019) explored the spatio-temporal correlations between the urban environment, traffic flow characteristics, and surge multipliers and proposed a general framework for predicting the short-term evolution of surge multipliers in real-time using a log-linear model with $\ell_1$-regularization, integrated with pattern clustering.  The modeling algorithm was validated by using Uber and Lyft data from Pittsburgh \cite{battifarano2019predicting}.

While there is a plethora of information available related to mobility and travel during this pandemic, it is critical to develop a robust methodological framework to accurately identify and analyze the key factors influencing human mobility and subsequently predict mobility using the selected factors in the time of such health crisis.  To this end, this study develops an analytical framework that helps determine the most significant factors affecting mobility by utilizing Granger causality followed by predicting mobility using linear regularization algorithms including the Ridge, and Least Absolute Shrinkage and Selection Operator (LASSO) modeling techniques.

The paper is organised as follows.  Section \ref{sec_granger} describes the theoretical concepts of Granger causality test, while Section \ref{sec_linear_model} explains the analytical methodology of ordinary least square method, and Ridge and LASSO linear regularization algorithms.  Additionally, Section \ref{sec_data} describes the time-series data analyzed in this study along with its descriptive statistics.  Moreover, Sections \ref{sec_causal_analysis} and \ref{sec_simulations} discuss the results of the causal analysis and the prediction of the regression models developed.  Finally, a summary and conclusion of this study along with its limitations and future research scope is included in Section \ref{sec_conclusions}.

\section{Granger Causality}\label{sec_granger}
In this section, we briefly discuss the concept of Granger causality with its two different methodological approaches.  Granger causality \cite{granger1969,granger1988} is a quantitative measure for inferring causal relationships between variables of a time-series data.  It is based on the following two principles.
\begin{assumption}\label{assumption_1}
Cause happens before the effect and the cause has unique information about the future of the effect.
\end{assumption}
The intuition behind the definition of Granger causality is as follows.  Suppose the goal is to predict the future of a variable $Y$.  If it happens that the prediction of $Y$ improves by considering the past of variables $Y$ and $X$ as opposed to considering the past of only $Y$, then it is said that $X$ ``Granger causes" $Y$. 
\begin{definition}
Under assumption \ref{assumption_1}, the hypothesis for testing Granger causality of $X$ on $Y$ is
\vspace{-4mm}
\begin{eqnarray}\label{granger_def}
P[Y(t+1)\in\Omega | I(t)]\neq P[Y(t+1)\in\Omega|I_{\not{X}}(t)]
\end{eqnarray}

where $P[Y(t+1)\in\Omega | I(t)]$ is the probability of $Y(t+1)$ belonging to the set $\Omega$ when the entire information till time $t$ is considered $(I(t))$ and $P[Y(t+1)\in\Omega|I_{\not{X}}]$ is the probability of $Y(t+1)$ belonging to the set $\Omega$ when $X$ is removed from the information set (denoted by $I_{\not{X}}(t)$).  When the above hypothesis (\ref{granger_def}) is satisfied, then we say $X$ Granger causes $Y$.
\end{definition}

\subsection{Bivariate Granger Causality}
Let $X_t$ and $Y_t$ be two time-series data, individually each of which can be represented by the following regressive models:
\vspace{-5mm}
\begin{eqnarray}\label{autoregressive_model}
\begin{aligned}
& X_t = \sum_{j=1}^\infty \alpha_{xj}X_{t-j} + \epsilon_{xt}, \quad \textnormal{var}(\epsilon_{xt})=\sigma_x\\
& Y_t = \sum_{j=1}^\infty \beta_{yj}Y_{t-j} + \epsilon_{yt}, \quad \textnormal{var}(\epsilon_{yt})=\sigma_y.
\end{aligned}
\end{eqnarray}

When considered jointly, the bivariate autoregressive models are:
\vspace{-2mm}
\begin{eqnarray}\label{bivariate_autoregressive_model}
\begin{aligned}
& X_t = \sum_{j=1}^\infty \alpha_{xj}'X_{t-j} + \sum_{j=1}^\infty\alpha_{xyj}'Y_{t-j}+\epsilon_{xt}'\\
& Y_t = \sum_{j=1}^\infty\beta_{yxj}'X_{t-j}+\sum_{j=1}^\infty \beta_{yj}'Y_{t-j} + \epsilon_{yt}',
\end{aligned}
\end{eqnarray}
where $\epsilon_{xt}'$ and $\epsilon_{yt}'$ are uncorrelated over time, such that their covariance matrix $\Sigma_{xy}$ is
\vspace{-2mm}
\begin{eqnarray}
\Sigma_{xy}' = \begin{pmatrix}
\sigma_{xx}' & \sigma_{xy}'\\
\sigma_{yx}' & \sigma_{yy}'
\end{pmatrix}.
\end{eqnarray}
With this, the Granger causality from each of the variable to the other is given by
\vspace{-2mm}
\begin{eqnarray}\label{granger_cause}
G_{X\to Y} = \ln \frac{\sigma_y}{\sigma_{yy}'};\quad G_{Y\to X} = \ln \frac{\sigma_x}{\sigma_{xx}'}.
\end{eqnarray}
Furthermore, the interdependence between the variables $X_t$ and $Y_t$ is given by
\vspace{-1mm}
\begin{eqnarray}
G_{X,Y} = \ln \frac{\sigma_x\sigma_y}{|\Sigma_{xy}'|},
\end{eqnarray}
where $|\cdot |$ denotes the determinant of a matrix.  Note that when $X_t$ and $Y_t$ are independent, $|\Sigma_{xy}'|=\sigma_{xx}'\sigma_{yy}'$ and hence $G_{X,Y}$ is zero.  When the interdependence is non-zero, in \cite{geweke}, it was shown that the total interdependence can be decomposed as
\vspace{-2mm}
\begin{eqnarray}
G_{X,Y} = G_{X\to Y} + G_{Y\to X} + G_{X\cdot Y},
\end{eqnarray}
where $G_{X\cdot Y}=\ln \frac{\sigma_{xx}'\sigma_{yy}'}{|\Sigma_{xy}'|}$ is the instantaneous causality between $X_t$ and $Y_t$.

\subsection{Conditional Granger Causality}

Equations (\ref{granger_cause}) give the Granger causality values for a bivariate time-series data.  However, in most real-life applications, the obtained time-series data has more than two variables.  In such cases, computing pairwise dependence using bivariate Granger causality may lead to ambiguous results \cite{geweke_conditional,chen2006frequency} because there may be direct or indirect causal links and in such circumstances, conditional Granger causality \cite{geweke_conditional} is more appropriate to infer causality.  For simplicity, the case with three variables is discussed and the general case with more variables follows directly.

Consider three time-series data $X_t$, $Y_t$ and $Z_t$ and suppose $Y_t$ has a pairwise causal influence on $X_t$.  We now consider the causal influence of $Y_t$ on $X_t$ that is mediated through $Z_t$.  Let the joint autoregressive model of $X_t$ and $Z_t$ be
\vspace{-4mm}
\begin{eqnarray}\label{bivariate_autoregressive_model_XZ}
\begin{aligned}
& X_t = \sum_{j=1}^\infty \alpha_{xj}X_{t-j} + \sum_{j=1}^\infty\alpha_{xzj}Z_{t-j}+\epsilon_{xt}\\
& Z_t = \sum_{j=1}^\infty\gamma_{zxj}X_{t-j}+\sum_{j=1}^\infty \gamma_{zj}Z_{t-j} + \epsilon_{zt},
\end{aligned}
\end{eqnarray}
with residual covariance matrix $\Sigma_{xz} = \begin{pmatrix}
\sigma_{xx} & \sigma_{xz}\\
\sigma_{zx} & \sigma_{zz}
\end{pmatrix}$.

Again, the joint autoregressive model for all the variables $X_t$, $Y_t$ and $Z_t$ is
\vspace{-2mm}
\begin{eqnarray}\label{bivariate_autoregressive_model_XYZ}
\begin{aligned}
& X_t = \sum_{j=1}^\infty \alpha_{xj}'X_{t-j} + \sum_{j=1}^\infty \alpha_{xyj}'Y_{t-j} +  \sum_{j=1}^\infty\alpha_{xzj}'Z_{t-j}+\epsilon_{xt}'\\
& Y_t = \sum_{j=1}^\infty \beta_{yxj}'X_{t-j} + \sum_{j=1}^\infty \beta_{yj}'Y_{t-j} +  \sum_{j=1}^\infty\beta_{yzj}'Z_{t-j}+\epsilon_{yt}'\\
& Z_t = \sum_{j=1}^\infty\gamma_{zxj}'X_{t-j} + \sum_{j=1}^\infty\gamma_{zyj}'Y_{t-j} + \sum_{j=1}^\infty \gamma_{zj}Z_{t-j} + \epsilon_{zt}',
\end{aligned}
\end{eqnarray}
with the residual covariance being 
\vspace{-3mm}
\begin{eqnarray}
\Sigma_{xyz}' = \begin{pmatrix}
\sigma_{xx}' & \sigma_{xy}' & \sigma_{xz}'\\
\sigma_{yx}' & \sigma_{yy}' & \sigma_{yz}'\\
\sigma_{zx}' & \sigma_{zy}' & \sigma_{zz}'\\
\end{pmatrix}.
\end{eqnarray}
With this, the Granger causality of $Y_t$ on $X_t$ conditioned on $Z_t$ is defined as
\begin{eqnarray}\label{conditional_granger_cause}
G_{Y\to X|Z} = \ln \frac{\sigma_{xx}}{\sigma_{xx}'}.
\end{eqnarray}

With this, if $G_{Y\to X|Z}>0$ and bivariate Granger analysis shows that $G_{Y\to X}\neq 0$, then the inclusion of $Y$ results in an improved prediction of $X$ and one can conclude that $Y$ influences $X$ directly.  If $G_{Y\to X|Z}=0$ and bivariate Granger causality analysis results in $G_{Y\to X}\neq 0$, then the influence of $Y_t$ on $X_t$ is entirely through $Z_t$.  

In the cases where there are more than three variables, conditional Granger causality can be defined similarly with the scalar $\sigma_{ij}$ and $\sigma_{ij}'$ replaced by corresponding elements of the residual covariance matrices.  However, it is to note that although Granger causality identifies the most influential variables, it does not provide the direction of the association between the explanatory and the dependent variables.

\section{Linear Models}\label{sec_linear_model}
In this section, we discuss the simple linear regression (ordinary least square or OLS) model and two of its variants, namely Ridge and LASSO regression models.

\subsection{Linear Regression}
Consider the data set $\{y_i,x_{i1},x_{i2},\cdots ,x_{iN}\}_{i=1}^n$ where $y_i$ is the $i^{th}$ observation of the dependent variable $y$ and $x_j$, $j=1,2,\cdots ,  N$ are the $N$ independent variables.  In case of linear regression, the model tries to fit a straight line by minimizing the residuals.  In particular, it assumes that the dependent variable can be expressed as a linear combination of the independent variables, as given by the following,
\vspace{-2mm}
\begin{eqnarray}\label{linear_regression}
y_i = \alpha_0 + \alpha_1x_{i1} + \alpha_2x_{i2} + \cdots + \alpha_Nx_{iN} + \epsilon_i, \quad i = 1, 2, \cdots , n,
\end{eqnarray}
where $\epsilon_i$ is the residual.

In matrix form, the equation (\ref{linear_regression}) can be written as
\vspace{-2mm}
\begin{eqnarray}
{\bf Y} = {\bf X}{\bf \alpha} + {\bf \epsilon},
\end{eqnarray}
where 
\vspace{-4mm}
\begin{eqnarray*}
{\bf Y} = [y_1,y_2,\cdots , y_n]^\top, {\bf X} =  \begin{pmatrix} 1 &  x_{11} & \cdots & x_{1N} \\
 1 & x_{21} & \cdots & x_{2N} \\
 \vdots & \vdots & \ddots & \vdots \\
 1 & x_{n1} & \cdots & x_{nN}
 \end{pmatrix}, {\bf \alpha} = [\alpha_0,\alpha_1,\cdots , \alpha_N]^\top, {\bf \epsilon} = [\epsilon_1,\epsilon_2,\cdots , \epsilon_n]^\top.
\end{eqnarray*}

The linear regression selects the parameters $\alpha_j$'s $(j = 0,\cdots , N)$ such that the norm of the residual for every $y_i$ $(i = 1,\cdots , n)$ is minimized.  Hence the optimal ${\bf \alpha}$ is obtained as a solution of the following optimization problem,
\vspace{-3mm}
\begin{eqnarray}\label{linear_regression_opt_prob}
\min_{\bf \alpha} \parallel {\bf Y} - {\bf X}{\bf \alpha}\parallel_2.
\end{eqnarray}
where $\parallel \cdot \parallel_2$ is the 2-norm of a vector.  The optimization problem (\ref{linear_regression_opt_prob}) is convex and can be solved efficiently either using convex optimization techniques or analytically, such that the optimal ${\bf \alpha}^\star$ is given by
\vspace{-1mm}
\begin{eqnarray}
{\bf \alpha}^\star = {\bf Y}{\bf X}^\dagger,
\end{eqnarray}
where ${\bf X}^\dagger$ is the Moore-Penrose inverse of $\bf X$.

\subsection{Ridge Regularization}
\begin{figure}[h!]
\centering
\subfigure[]{\includegraphics[scale=.28]{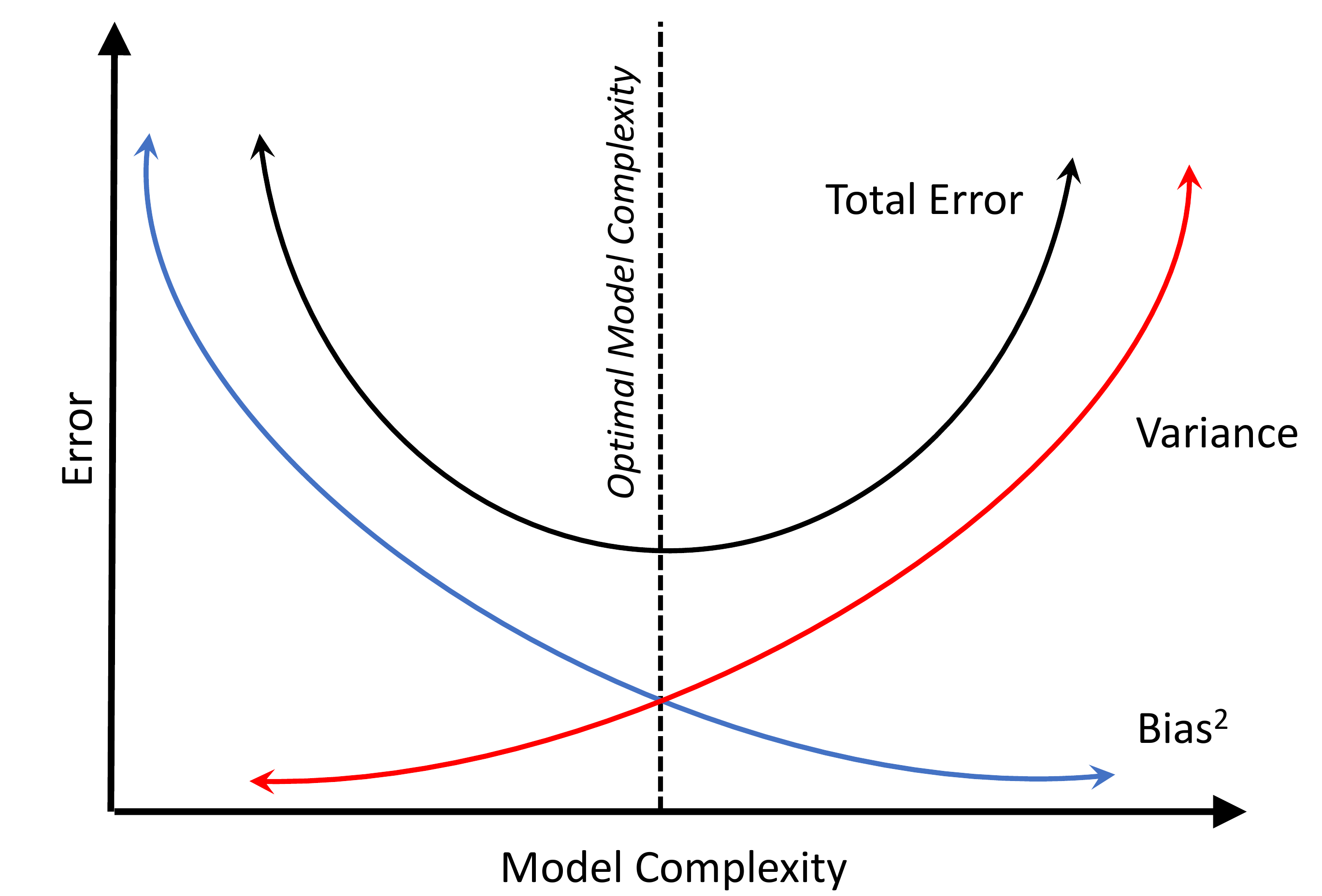}}
\subfigure[]{\includegraphics[scale=.4]{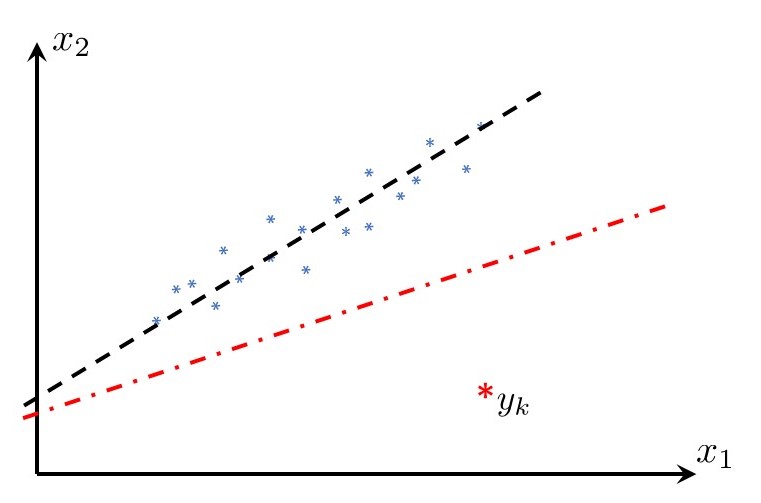}}
\caption{(a) Bias-variance trade-off. (b) Effect of outlier on linear regression.}\label{bias_variance}
\end{figure}

A major drawback of linear regression is that this algorithm has low bias and high variance \cite{mitchell_book}.  This means that the linear regression may perform well on the train data, but it may not generalize well to the test data set, thereby making the model performance unsatisfactory.  In machine learning literature, this phenomenon is known as Bias-Variance trade-off \cite{mitchell_book}.  The intuition of bias-variance trade-off is explained in Figure \ref{bias_variance}(a).  Usually, with a highly complex model, it is possible to fit the training data as closely as possible.  In this case, the training error is zero and the model is said to have low bias.  However, the highly complex model may not generalize well to the test data, thus making the test error large.  This is due to the overfitting of the training data.  The complex model, which overfits the training data and produces high test error, is said to have high variance.  This situation is often reversed if the model considered is fairly simple.  Ridge regression, which puts a 2-norm ($\ell_2$-norm) constraint on the set of coefficients, is able to overcome this challenge efficiently. 

Another drawback of OLS is that the obtained model is highly influenced by the outliers in the training data set.  For example, as in Figure \ref{bias_variance}(b), the outlier data point $y_k$ results in the linear model fit as represented by the red line.  However, it is obvious by looking at the overall data that the model fit depicted by the black line is the more appropriate linear fit to the data.

Additionally, on many occasions, the real-life data is noisy or uncertain.  When the OLS attempts to fit that noise in the data, it eventually results in overfitting and consequently degrading its performance for model prediction.  As stated earlier, the data for this study were obtained mostly from smartphone devices, and the chances of acquiring this data may also be subject to individual user's discretion, so it is reasonable to assume that the data utilized in this study may contain some noise or uncertainty in it.  To account for the noise in the data, it is assumed that there is some uncertainty, $\Delta \Y$ and $\Delta \X$, in the dependent and independent variables, respectively.  It is assumed that the uncertainties in both $\Y$ and $\X$ are bounded, i.e. there exists some positive real number $\rho>0$ such that  $\parallel \Delta \X\parallel_2\leq \rho$ and $\parallel \Delta \Y \parallel_2\leq \rho$.  With this, the optimization problem (\ref{linear_regression_opt_prob}) is modified to a min-max optimization problem \cite{robust_koopman_acc,robust_koopman_journal} given by,
\vspace{-1mm}
\begin{eqnarray}\label{ridge_uncertain}
\min_{\A}\max_{\substack{\parallel \Delta \X\parallel_2\leq \rho,\\ \parallel \Delta \Y\parallel_2\leq \rho}} \parallel (\Y + \Delta \Y) - (\X + \Delta X)\A\parallel_2.
\end{eqnarray}

Min-max optimization problems are generally hard to solve, but in this case, the optimization problem (\ref{ridge_uncertain}) can be equivalently expressed as a convex optimization problem as follows,

\begin{theorem}
The optimization problem  
\vspace{-5mm}
\begin{eqnarray}
\min_{\A}\max_{\substack{\parallel \Delta \X\parallel_2\leq \rho,\\ \parallel \Delta \Y\parallel_2\leq \rho}} \parallel (\Y + \Delta \Y) - (\X + \Delta X)\A\parallel_2.
\end{eqnarray}
is equivalent to the following,
\vspace{-8.5mm}
\begin{eqnarray}\label{ridge_opt}
\min_{\A} \parallel \Y - \X\A\parallel_2 + \lambda_2\parallel \A\parallel_2,
\end{eqnarray}
where $\lambda_2$ is a positive real number, depending on the uncertainty bound $\rho$.
\end{theorem}
\textbf{Proof.} For proof, see \cite{robust_koopman_acc,robust_koopman_journal}.
$\hfill\square$

The optimization problem (\ref{ridge_opt}), known as Ridge regression, is a convex problem and can be solved efficiently using any of the available convex optimization problem solvers.  The parameter $\lambda_2$ is called the regularization parameter and it acts as a trade-off between the OLS cost and the cost on the coefficients $\A$.

\subsection{LASSO Regularization}
Apart from the 2-norm regularization, another popular regularization is the $\ell_1$-norm regularization, which is also known as Least Absolute Shrinkage and Selection Operator (LASSO) regression.  In particular, instead of putting a 2-norm cost on the coefficients of the linear model, LASSO employs a 1-norm cost on the coefficients. Hence, the LASSO regression model is obtained by solving the following optimization problem:

\vspace{-4mm}
\begin{eqnarray}\label{lasso_opt_prob}
\begin{aligned}
\min_{\bf \alpha}\qquad & \parallel {\bf Y} - {\bf X}{\bf \alpha}\parallel_2 + \lambda_1 \parallel \A \parallel_1\\
\textnormal{subject to}\quad &\parallel \A \parallel_1\leq t,
\end{aligned}
\end{eqnarray}
where $\parallel \cdot \parallel_1$ is the $1$-norm of a vector and the bound $t$ is the tuning parameter.  If $t$ is large, it has no effect on the regression coefficients $\alpha_i$s and in this case, the solution to the optimization problem (\ref{lasso_opt_prob}) approach the solution of normal linear regression optimization problem (\ref{linear_regression_opt_prob}) in the limit of large $t$.  However, when the bound $t$ is small, the parameters $\alpha_i$s are constrained and hence are shrunk and are smaller versions of the original least squares estimates.  The $1$-norm minimization puts constraints on parameters that shrink coefficients towards zero and thus leads to a sparse solution for the linear model.

\section{Data Description}\label{sec_data}
Data for this study were collected and combined from multiple web-based open-access sources.  The majority of data used in this analysis were requested and obtained from the COVID-19 Impact Analysis Platform developed at the Maryland Transportation Institute of the University of Maryland (UMD) \cite{UMDCOVID:2020}.  This platform provides both state and county-based information for 50 states in the U.S. and the District of Columbia.  To match with the data available from other sources, for the purpose of this study, state-wise daily time-series data were requested from this platform.  The relevant statewide data obtained and analyzed from this source include the daily number of new COVID-19 confirmed cases per 1,000 people, social distancing index, percent of out of county and state trips, transit mode share, population, percent of people older than 60 years, percent of African American or Hispanic Americans, median income, percent of male population, number of hot spots per 1,000 people, unemployment rate, percent of people working from home, among others.  Social distancing index in the data set indicates the increasing space between individuals and decreasing frequency of contact and is represented as an integer from 0 to 100, where 0 indicates no social distancing in the state and 100 indicates all residents are staying at home. 

\begin{figure}[htp!]
\centering
\includegraphics[scale=0.6]{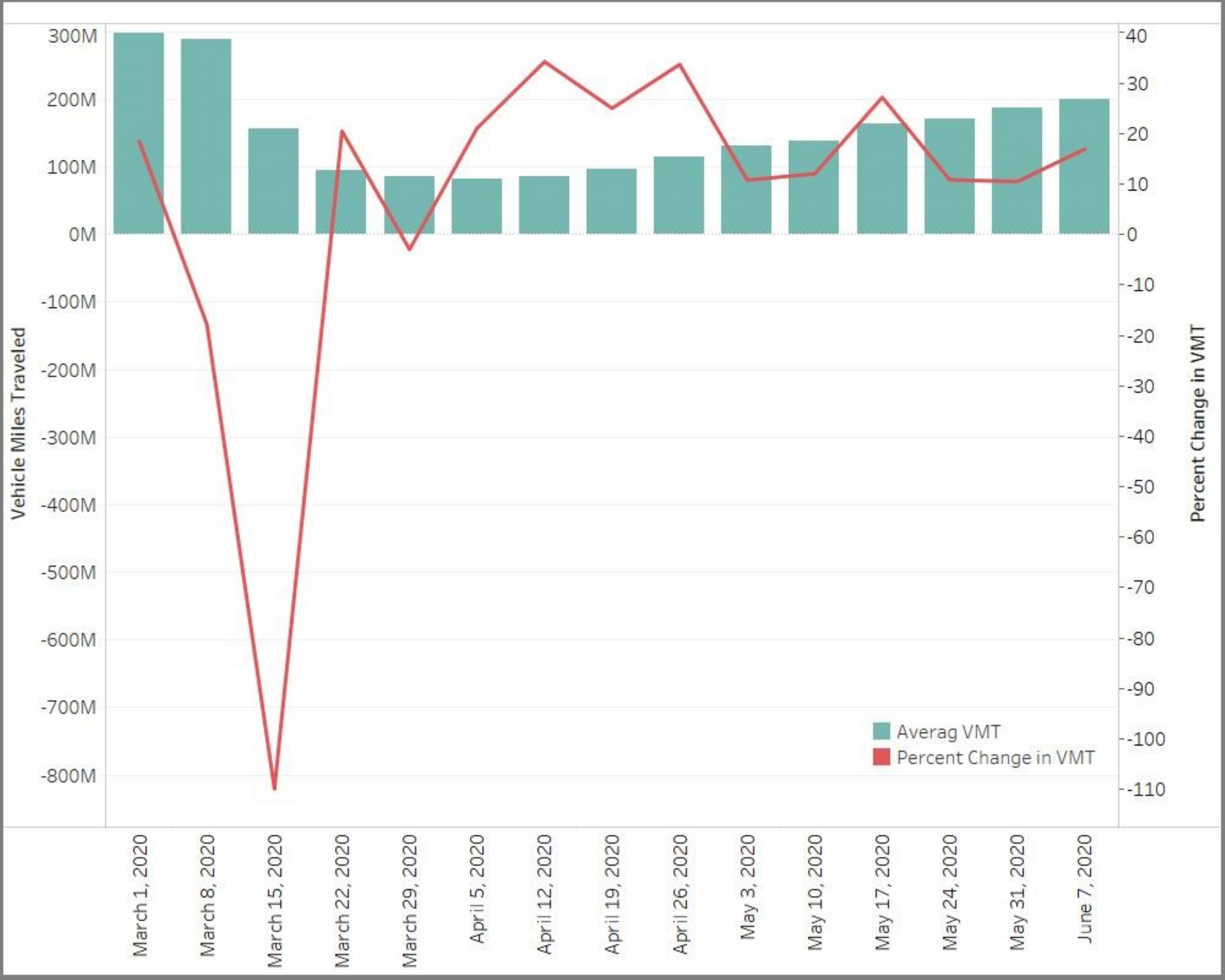}
\caption{The average vehicle miles traveled during the analysis period.}\label{Vehicle_miles_traveled}
\end{figure}

Additional information was collected and appended with the data obtained from the UMD platform.  The data for daily vehicle miles traveled (VMT) starting from March 1, 2020 through June 19, 2020 for 48 states (excluding Alaska and Hawaii), and the District of Columbia were requested and obtained from the Streetlight Data \cite{VMT}.  The movement trends over time by geography, across different categories of places such as retail and recreation, groceries and pharmacies, parks, transit stations, workplaces, and residential, reported by Google, were further joined with the study data set \cite{googlemobility}.  Additionally, the percent of the population at different education levels and gross domestic product (GDP) information were collected from the U.S. Census Bureau \cite{USCensus}.  Information on the percent of population staying at home during this time was obtained from the Bureau of Transportation Statistics \cite{BTS}.    Furthermore, several states have exercised travel restrictions in the form of stay at home order, limitations on gatherings, domestic travel limitations, or school closures.  This information was obtained from the COVID-19 State and Territory Action Tracker \cite{ActionTracker} provided by ESRI (Environmental Systems Research Institute) \cite{ESRI}. 
 
{\tiny
\begin{table}[H]
\centering
\caption{Descriptive Statistics of the Data}\label{independent_variables_table}
    \begin{tabular}{|p{8cm}|c|c|c|c|c|c|}
\hline
	\textbf{Parameter} & \textbf{Minimum} & \textbf{Maximum} & \textbf{Mean} & \textbf{Std. Dev} \\ \hline \hline
	Daily vehicle miles traveled ($10^{6}$ miles) & 0.36 & 2,519.35 & 152.48 & 200.31 \\ \hline
	Number of new COVID-19 cases (per 1,000 people) & 0 & 0.72 & 0.06 & 0.07 \\ \hline
	Social distancing index & 10 & 83 & 38.05 & 16.07 \\ \hline
	Percent of out of county trips & 6.9 & 52.1 & 28.01 & 7.17 \\ \hline
	Percent of out of state trips & 0.4 & 52.1 & 6.8 & 6.51 \\ \hline
	Total population & 577,737 & 39,557,045 & 6,632,847 & 7,336,262 \\ \hline
	Transit mode share & 0.29 & 34.83 & 3.66 & 6.34 \\ \hline
	Percent of population older than 60 years & 15 & 27 & 21.94 & 2.24 \\ \hline
    Median income (US dollars) & 44,445 & 84,342 & 61,137 & 10,219.58 \\ \hline
    Percent of Hispanic or African American population & 1.5 & 48.5 & 11.85 & 10.32 \\ \hline
    Percent of male & 47.46 & 51.34 & 49.32 & 0.73 \\ \hline
    Number of hot spots (per 1,000 people) & 47.46 & 51.34 & 49.32 & 0.73 \\ \hline   
	Unemployment rate & 2.2 & 46.6 & 16.67 & 9.29 \\ \hline
	Percent of population working from home & 2.3 & 55.7 & 22.88 & 9.8 \\ \hline
    Number of tests conducted (per 1,000 people) & 0 & 181.14 & 22.07 & 25.63 \\ \hline 	
	Change in the number of trips to transit stations from baseline & -81 & 44 & -27.32 & 22.55 \\ \hline
	Change in the number of trips to retail and recreation places from baseline & -77 & 24 & -23.9 & 19.15 \\ \hline
	Change in the number of trips to grocery and pharmacy from baseline & -62 & 48 & -2.84 & 13.01 \\ \hline	
	Change in the number of trips to parks from baseline & -77 & 388 & 38.184 & 60.2 \\ \hline	
	Change in the number of trips to workplaces from baseline & -78 & 18 & -32.08 & 17.55 \\ \hline	
	Change in the number of trips to residential places from baseline & -5 & 33 & 11.87 & 7.31 \\ \hline
    Population staying at home & 108,677 & 14,180,383 & 1,646,749 & 1,932,225 \\ \hline	
    Socioeconomic status & 0 & 1 & 0.51 & 0.29 \\ \hline	
	State employee travel restriction & 0 & 1 & 0.71 & 0.45 \\ \hline  
    Stay-at-home order or guidance & 0 & 1 & 0.76 & 0.43 \\ \hline	
    School closures & 0 & 1 & 0.96 & 0.20 \\ \hline
	Closure of some or all facilities & 0 & 1 & 0.86 & 0.35 \\ \hline
	Mandatory or recommended domestic travel limitations & 0 & 1 & 0.43 & 0.49 \\ \hline	
	Mandatory statewide mask policy & 0 & 1 & 0.69 & 0.46 \\ \hline	
	Median age & 31 & 45.1 & 38.66 & 2.38 \\ \hline
	GDP in 2019 ($10^{9}$ U.S. dollars) & 30.5 & 2,792.03 & 382.32 & 488.51 \\ \hline
	Percent of population aged 25 years or over with high school or higher & 83.8 & 93.9 & 89.74 & 2.58 \\ \hline

\end{tabular}
\end{table}
}

Following the joining of the data from various aforementioned sources, a thorough screening and quality check of the data was performed for any missing values.  The final data set includes daily data starting from March 1, 2020 through June 13, 2020 from 48 U.S. states excluding Alaska and Hawaii and including the District of Columbia, and consists of a total of 5,145 observations for further analysis.  In the final data set, the two states, Alaska and Hawaii, were excluded as the VMT data was not available for the same.

The preliminary analysis examined several predictor variables of interest that were initially included in the analysis.  The distribution of the dependent variable i.e., the daily VMT during the analysis period (from March 1, 2020 through June 13, 2020) is shown in Figure \ref{Vehicle_miles_traveled}.  As can be clearly seen from Figure \ref{Vehicle_miles_traveled}, the average VMT dropped substantially from the beginning of March, 2020, around the time when the U.S. started experiencing rapid community outbreaks of the virus and the country declared a national emergency, and it continued to remain low until around end of April, 2020.  However, it is interesting to see that the daily VMT gradually increased from around early May, 2020, although the number of daily COVID-19 cases continued to grow considerably over time.

 As stated earlier, the compiled data set includes a large number of variables of interest that may impact the daily VMT.  Table \ref{independent_variables_table} presents the descriptive statistics, including the minimums, maximums, means, and standard deviations of thirty-one variables that were initially considered as the set of potential determinants.  For the purpose of scaling the data, VMT, population, median income, and GDP were included in their natural log forms.

\section{Causal Analysis of Data}\label{sec_causal_analysis}

In any multivariate time-series data, there is causal influence between the variables involved.  Correlations between the variables quantify the extent of their interdependence, but they do not specifically identify the cause and the effect.  In particular, the correlations are symmetric in the variables and lack a directional sense.  However, to better understand the relationships among the variables in the data, identification of the causal structure and influential predictor variables is critical.  For example, for a large time-series data with a massive number of predictor variables, it is advisable to consider a reduced order model, which can be done by identifying the most influential variables.  In this study, the goal of determining the most important predictor variables from the large set of time-series data was accomplished by employing the Granger causality test. 

\begin{figure}[htp!]
\centering
\includegraphics[scale=0.38]{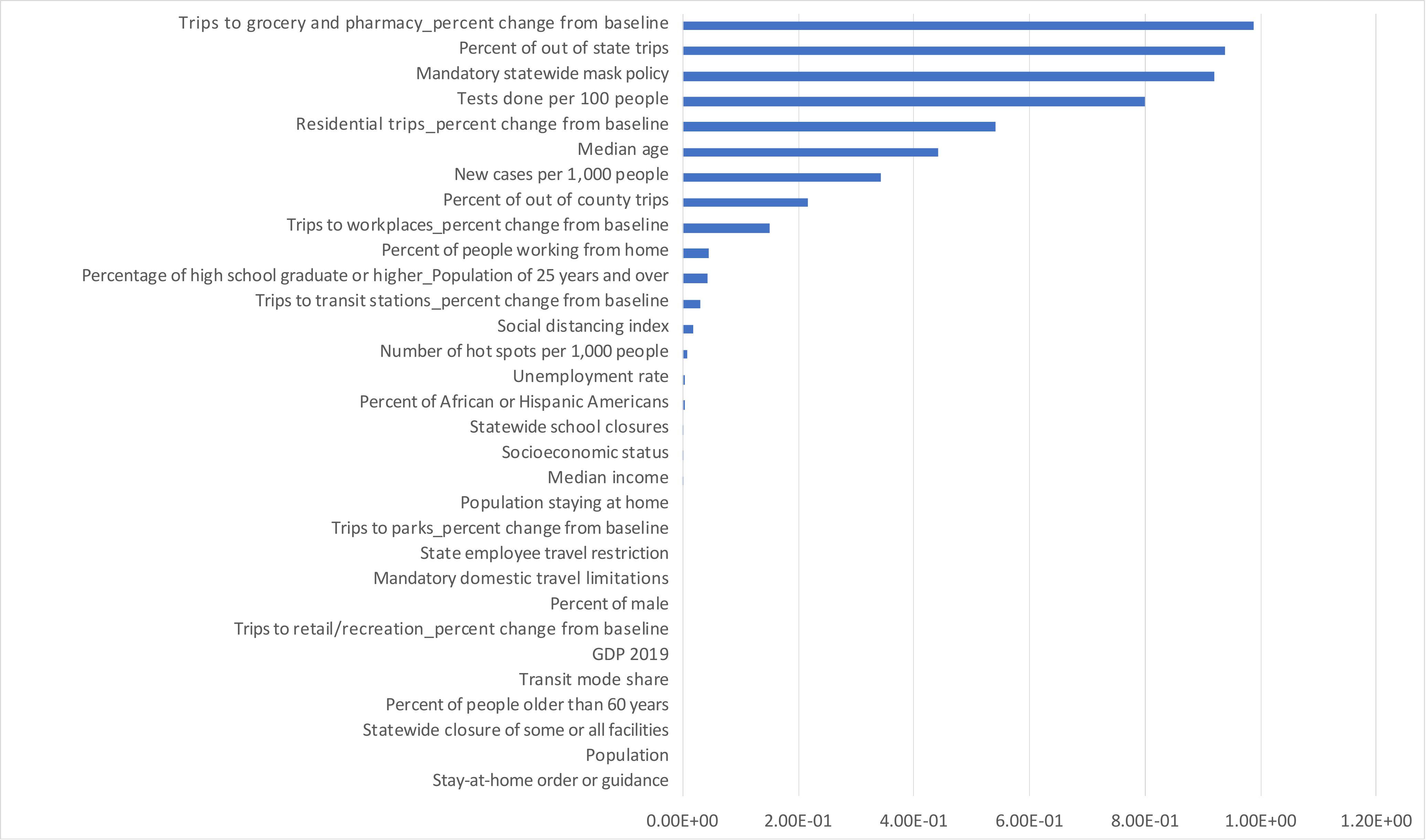}
\caption{Bivariate Granger causality test values of the predictor variables to dependent variable.}\label{bivariate granger}
\end{figure}

As discussed earlier in section \ref{sec_granger}, Granger causality between two variables can be analyzed by either doing a pairwise study by developing bivariate VAR model or by conditional Granger causality by considering multivariate VAR model.  For multivariate data, bivariate Granger causality analysis often provides ambiguous results \cite{geweke_conditional,chen2006frequency}.  For validation purposes, the study initially carries out a bivarite Granger causality test whose results are shown in Figure \ref{bivariate granger}.  In this figure, the Granger causality test values of all predictor variables on the dependent variable (daily VMT) using bivariate models are displayed.  The lags in this VAR model were chosen by Akaike Information Criterion (AIC) and the optimal lag was found to be 4 for the bivariate VAR models.  From Figure \ref{bivariate granger}, it can be seen that the model was not efficient in determining causality for most of the variables and several apparently important predictors were not correctly captured in explaining the causal relationships between the explanatory and dependent variable.  For example, the most influential predictor variable was identified as \emph{the change in the percent of trips to grocery and pharmacy from the average baseline}, whereas factors like \emph{social distancing index}, or \emph{population staying at home} or \emph{percent of people working from home} etc. had very little to no causal influence on the dependent variable.  This observation is counter-intuitive and supports the previous theoretical argument that bivariate Granger causality tests provide incoherent results for multivariate time-series data. 

\begin{figure}[htp!]
\centering
\includegraphics[scale=0.38]{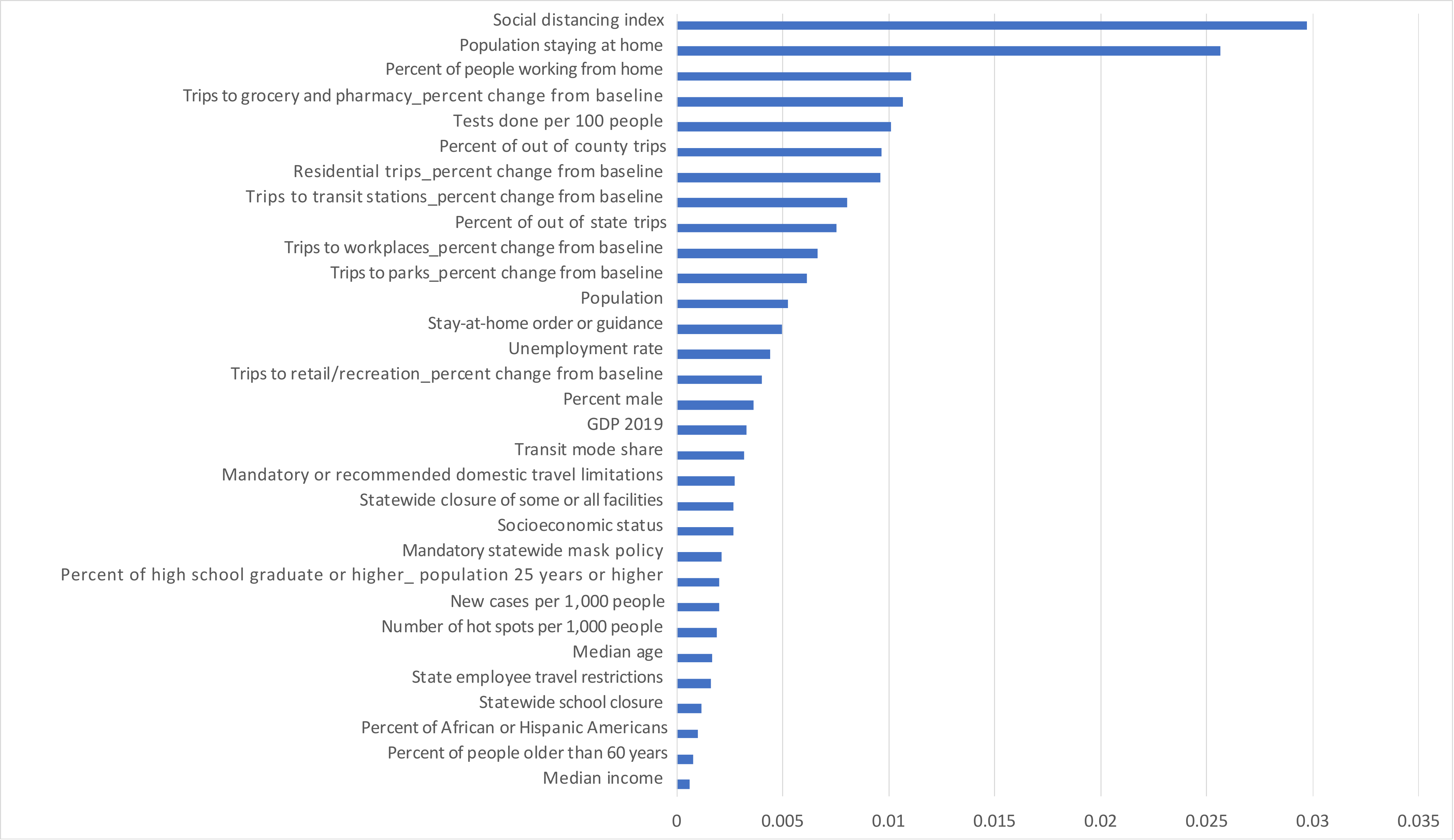}
\caption{Multivariate Granger causality test values of the predictor variables to dependent variable.}\label{multivariate granger}
\end{figure}

For conditional Granger causality test, a multivariate VAR model was developed and similar to the bivariate model, the optimal lag for the VAR model was determined by the Akaike Information Criterion and this optimal lag was obtained as 10.  While determining the conditional Granger causality (\ref{conditional_granger_cause}) of the $i^{th}$ predictor variable on the dependent variable, the conditional set was chosen as all the predictor variables except the $i^{th}$ predictor variable.  Figure \ref{multivariate granger} presents the results of the conditional Granger causality test by showing the causality values for each of the predictor variables.  As can be seen from Figure \ref{multivariate granger}, the conditional Granger causality efficiently identified the important predictors from a pool of a large number of explanatory variables.  For example, unlike the bivariate causality test, \emph{social distancing index} or \emph{population staying at home} were identified as the important variables in explaining the daily vehicle miles traveled.

\section{Regression Models and Prediction}\label{sec_simulations}

In the previous section, the influence of the predictor variables on the VMT was computed and was rank-ordered according to the influence each predictor variable has on the VMT.  In this section, the most influential predictor variables identified by Granger causality test were used to develop three different reduced order linear models, namely,  ordinary least squares, LASSO, and Ridge models.  In particular, the largest value of the causal influence was around 0.03 and the cut-off was selected to be 0.003.  This cut-off resulted in the selection of seventeen predictor variables and with these selected variables, reduced order models for predicting VMT were developed.

\subsection{Model Training}
The final data set used in this study included information from March 1, 2020 through June 13, 2020.  After splitting the data set into two parts for training and testing, the train and test data sets ranged from March 1, 2020 through May 18, 2020 and from May 19, 2020 through June 13, 2020, respectively.  The first (train) part was utilized for training the regression models, while the latter part (test) was used to test the efficiency of the prediction of the obtained regression models.  As mentioned earlier, the seventeen most influential variables were considered for the reduced order models.  It is important to note that, except for the OLS model, the other computation of Ridge and LASSO models (optimization problems (\ref{ridge_opt}) and (\ref{lasso_opt_prob})) involve one regularization parameter and in all the optimization problems the $\lambda_i$s were chosen such that, $0<\lambda_i\leq 1$.  

\subsubsection{Interpretation and Discussion of the Model Results}
Table \ref{independent_variable_reduced_table} compares the optimal coefficients of the final set of explanatory variables between all three regression modeling techniques.  As can be seen from Table \ref{independent_variable_reduced_table}, the coefficients of the predictors were fairly comparable across all modeling techniques.  As expected, when the number of new COVID cases per 1,000 people increased, the daily vehicle miles traveled decreased.  Expectedly, with the increase in the social distancing index, percent of people working from home, statewide closure, and tests done per 100 people, the daily VMT decreased.  Additionally, the vehicle miles traveled per day increased with the increase in population, unemployment rate, person of out of county trips, socio-economic status, and the increase in percent of trips to transit stations, retail and recreational places, grocery and pharmacy, workplaces, and residences.  The predictor for population staying at home was rightly captured by regularization methods, namely LASSO and Ridge regressions, and showed a negative association with daily VMT.  Lastly, although the negative association between the increase in the percent of trips to parks and daily VMT seemed counter-intuitive,  this could be partially due to the reason that people might also rely on non-motorized transport to go to parks and such trips would not be captured by the daily VMT.

\begin{figure}[htp!]
\centering
\includegraphics[scale=0.45]{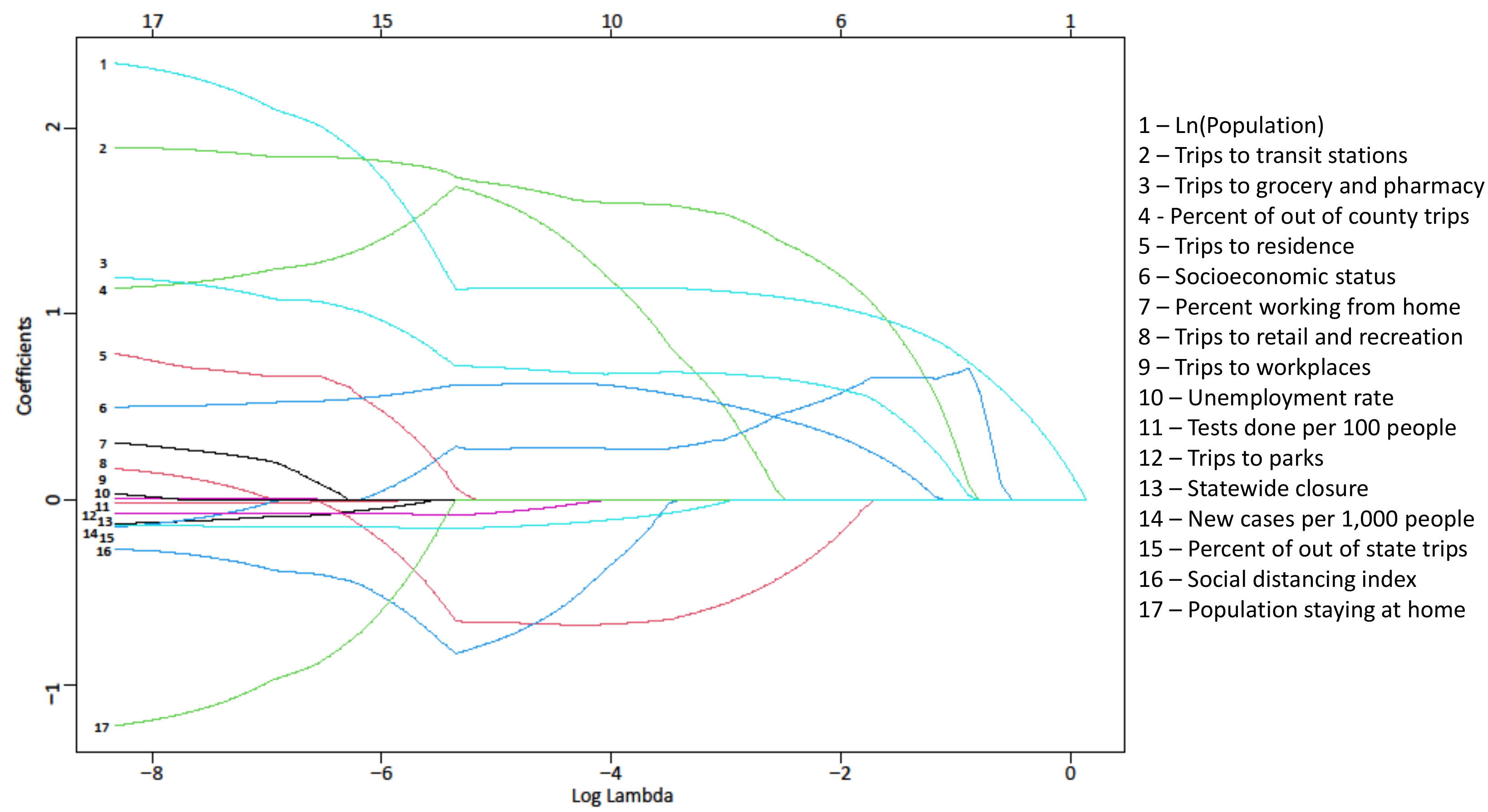}
\caption{Relative importance of the predictors in terms of log-lambda for LASSO regression.}\label{log_lambda_lasso}
\end{figure}

\begin{figure}[htp!]
\centering
\includegraphics[scale=0.45]{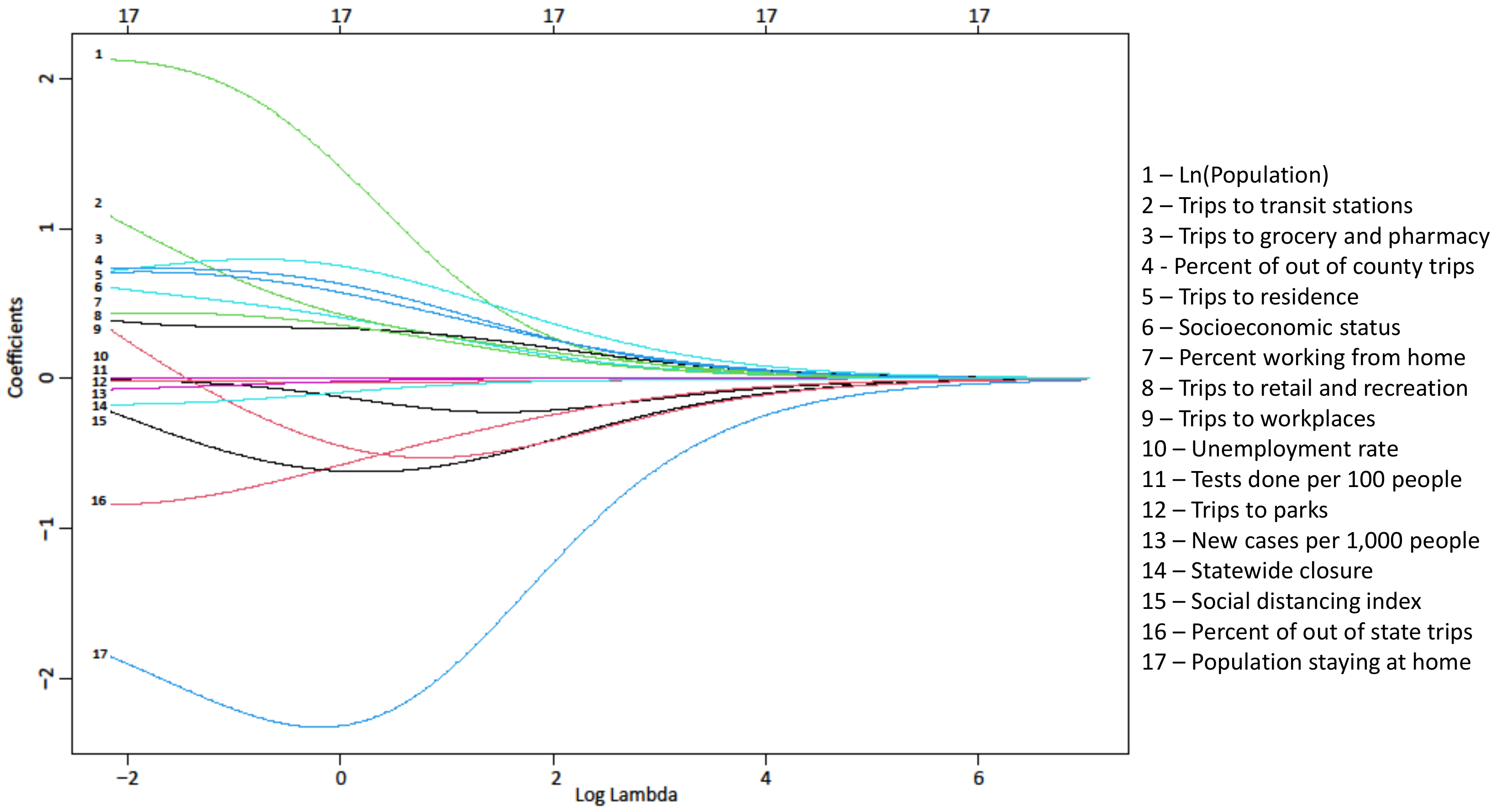}
\caption{Relative importance of the predictors in terms of log-lambda for Ridge regression.}\label{log_lambda_ridge}
\end{figure}

Furthermore, the log-lambda plots shown in Figures \ref{log_lambda_lasso} and \ref{log_lambda_ridge} depicts how the predictor variables that enter in the model, varied across the LASSO and Ridge regression techniques as the regularization parameter change.  When the regularization parameter $\lambda$ is small, the contribution of the regularization part to the cost functions in the optimization problems (\ref{ridge_opt}) and (\ref{lasso_opt_prob})  is small and as such all these optimization problems are reduced to the simple linear regression or OLS.  However, as $\lambda$ is increased, the weight on the regularization component in the optimization problems increases and as such, the coefficients of the independent variables become smaller and approach zero.  It is clearly seen from Figures \ref{log_lambda_lasso} and \ref{log_lambda_ridge} that all the coefficients did not approach zero at the same time.  In particular, the important variables remained non-zero for larger values of $\lambda$ as compared to the relatively non-important variables.  Based on the order of importance, the coefficients of the explanatory variables differed slightly between the LASSO and Ridge regularization techniques.  This difference could be due to the characteristic of the Ridge regularization that reduces the norm of the coefficients more uniformly, while the LASSO model attempts to set as many coefficients to zero as possible. 

\begin{table}[H]
\centering
\caption{Comparison of Model Coefficients between the Linear, LASSO, and Ridge Regressions}\label{independent_variable_reduced_table}
\begin{tabular}{|p{10cm}|c|c|c|c|}
\hline
		\textbf{Parameter} & 	\textbf{Linear} & 	\textbf{LASSO} & 	\textbf{Ridge} \\ \hline \hline
	Intercept & 3.119 & -1.319 & -1.049  \\ \hline
	New COVID-19 cases per 1,000 people & -0.010 & -0.148 & -0.128  \\ \hline
	Social distancing index & -0.844 & -0.266 & -0.169  \\ \hline
	Percent of out of county trips & 2.126 & 1.064 & 1.139  \\ \hline
	Percent of out of state trips & -1.854 & -0.177 & -0.263  \\ \hline
	Ln(Population) & 0.607 & 2.504 & 2.351  \\ \hline
	Unemployment rate & 0.005 & 0.005 & 0.005  \\ \hline
	Percent of population working from home & -0.221 & 0.356 & 0.307 \\ \hline
	Tests done per 100 people & -0.015 & -0.021 & -0.019 \\ \hline	
	Trips to transit stations-percent change from baseline & 1.081 & 1.920 & 1.899 \\ \hline
	Trips to retail and recreation places-percent change from baseline & 0.707 & 0.225 & 0.145 \\ \hline	
	Trips to grocery and pharmacy-percent change from baseline & 0.717 & 1.265 & 1.199 \\ \hline	
	Trips to parks-percent change from baseline & -0.073 & -0.066 & -0.069 \\ \hline	
	Trips to workplaces-percent change from baseline & 0.387 & 0.046 & 0.033 \\ \hline
	Trips to residence-percent change from baseline & 0.327 & 0.847 & 0.788 \\ \hline
	Population staying at home & 0.432 & -1.368 & -1.216  \\ \hline 
	Socioeconomic status & 0.731 & 0.486 & 0.501  \\ \hline
	Statewide closure of some or all facilities & -0.175 & -0.136 & -0.139  \\ \hline
\end{tabular}
\end{table}

\newpage

\subsection{Prediction Performance}

Following the development of the models using the training data, the prediction of the dependent variable (daily VMT) was tested and compared between the three regression techniques.  Essentially, the performance of the three different models on the test data set was evaluated, and the predicted values were compared with the observed values.  Ultimately, the root mean square errors (RMSEs) from all models provided the measure of performance and efficiency of the models.

\begin{table}[htp!]
\centering
\caption{Root mean square error (RMSE) results of training and test data}\label{RMSE_table}
\begin{tabular}{|c|c|c|c|}
\hline
		\textbf{Modeling Technique} & 	\textbf{ RMSE in Training} & 	\textbf{RMSE in Test} \\ [12 pt]\hline \hline
	Linear & 0.3733 & 0.4330 \\[10 pt] \hline
	LASSO & 0.3346 & 0.3998 \\ [10 pt] \hline
	Ridge & 0.3348 & 0.3952 \\ [10 pt] \hline
\end{tabular}
\end{table}

The RMSEs of the three different models utilizing both the train and test data set are presented in Table \ref{RMSE_table}.  The comparison of the RMSEs between the three modeling techniques clearly shows that the Ridge regression performed the best for both train and test data by having the least RMSE among all models.  This is expected, as the Ridge regression provides superior prediction by overcoming the issue of overfitting with low variance and better generalization to the test data compared to other regularization methods (refer to Figure \ref{bias_variance}).  Additionally, based on the RMSEs, also the LASSO model provided better prediction performance compared to the OLS.  However, the higher error in the test data in all models could partially be due a much smaller sample size and the consistent gradual increase in VMT in the test period as opposed to the change in VMT from the average baseline in both directions during the training data period. 

\begin{figure}[htp!]
\centering
\subfigure[]{\includegraphics[scale=.37]{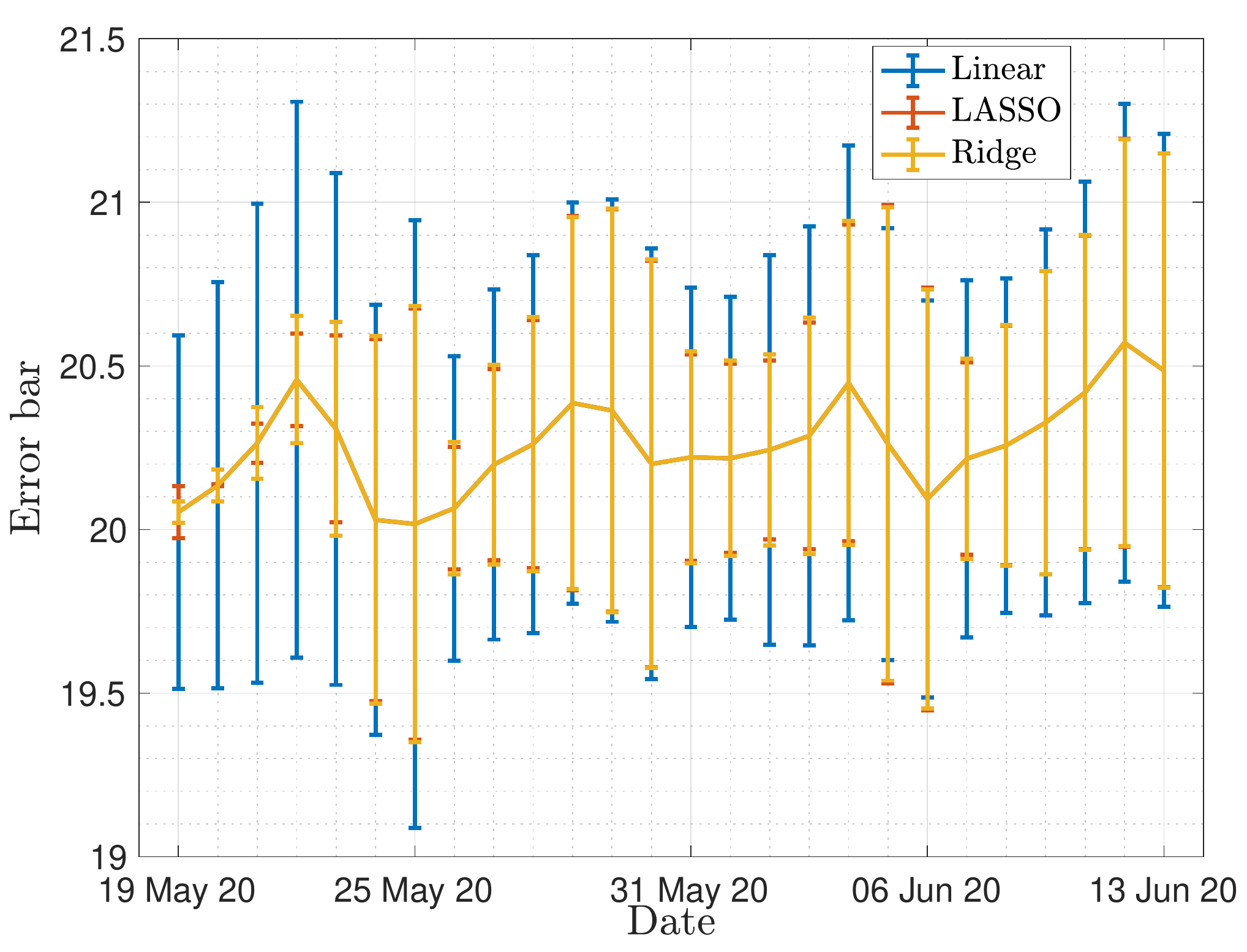}}
\subfigure[]{\includegraphics[scale=.37]{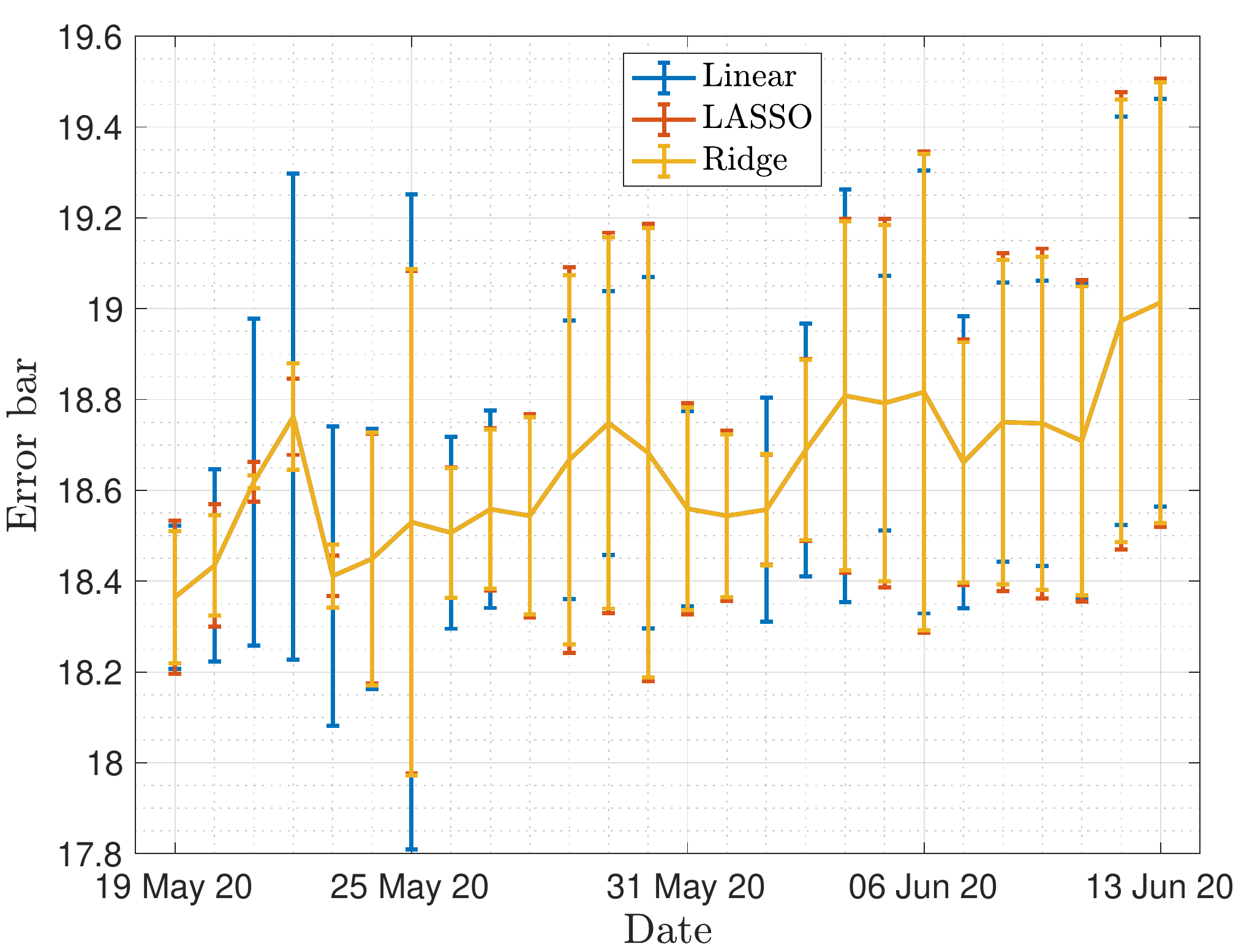}}
\subfigure[]{\includegraphics[scale=.37]{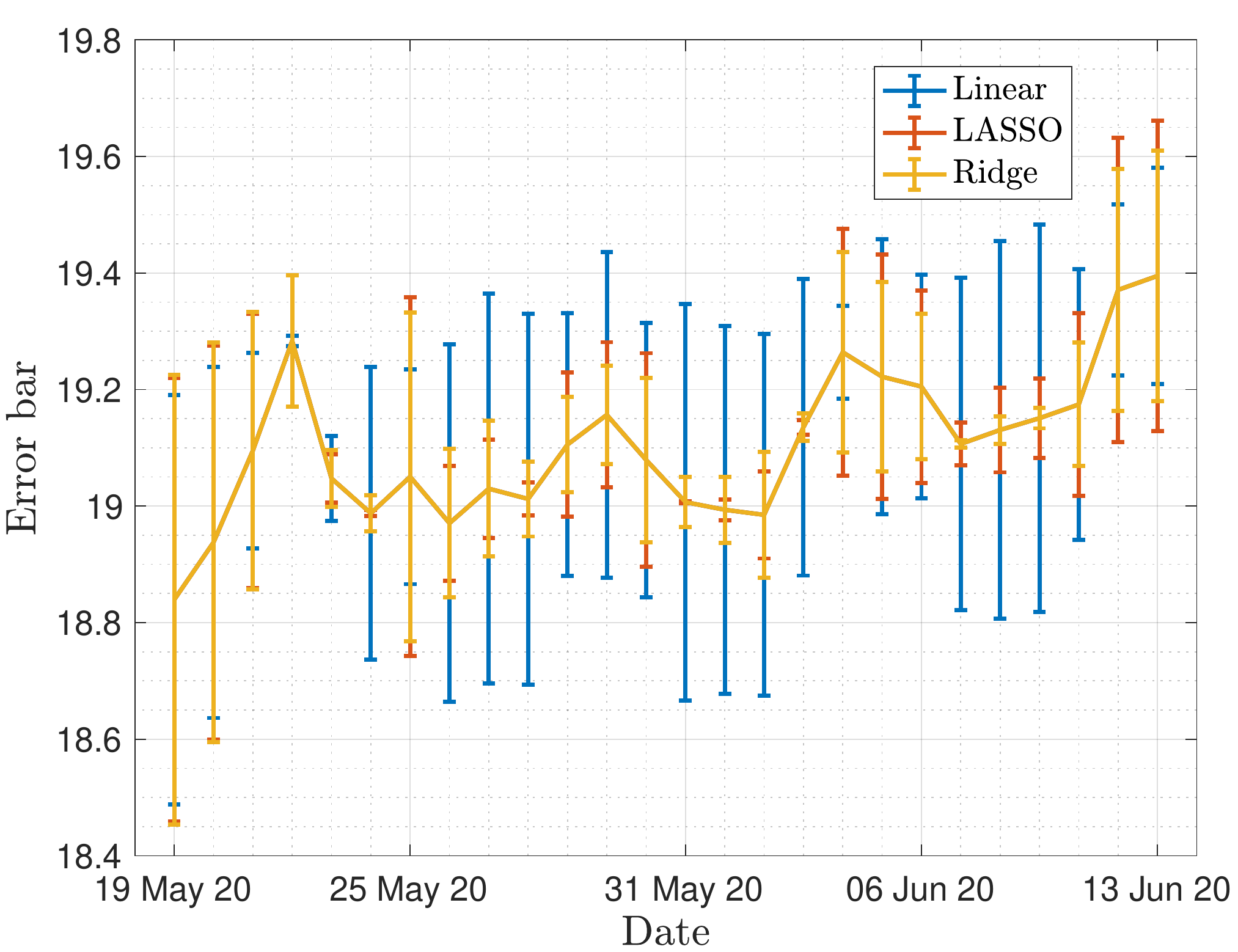}}
\subfigure[]{\includegraphics[scale=.37]{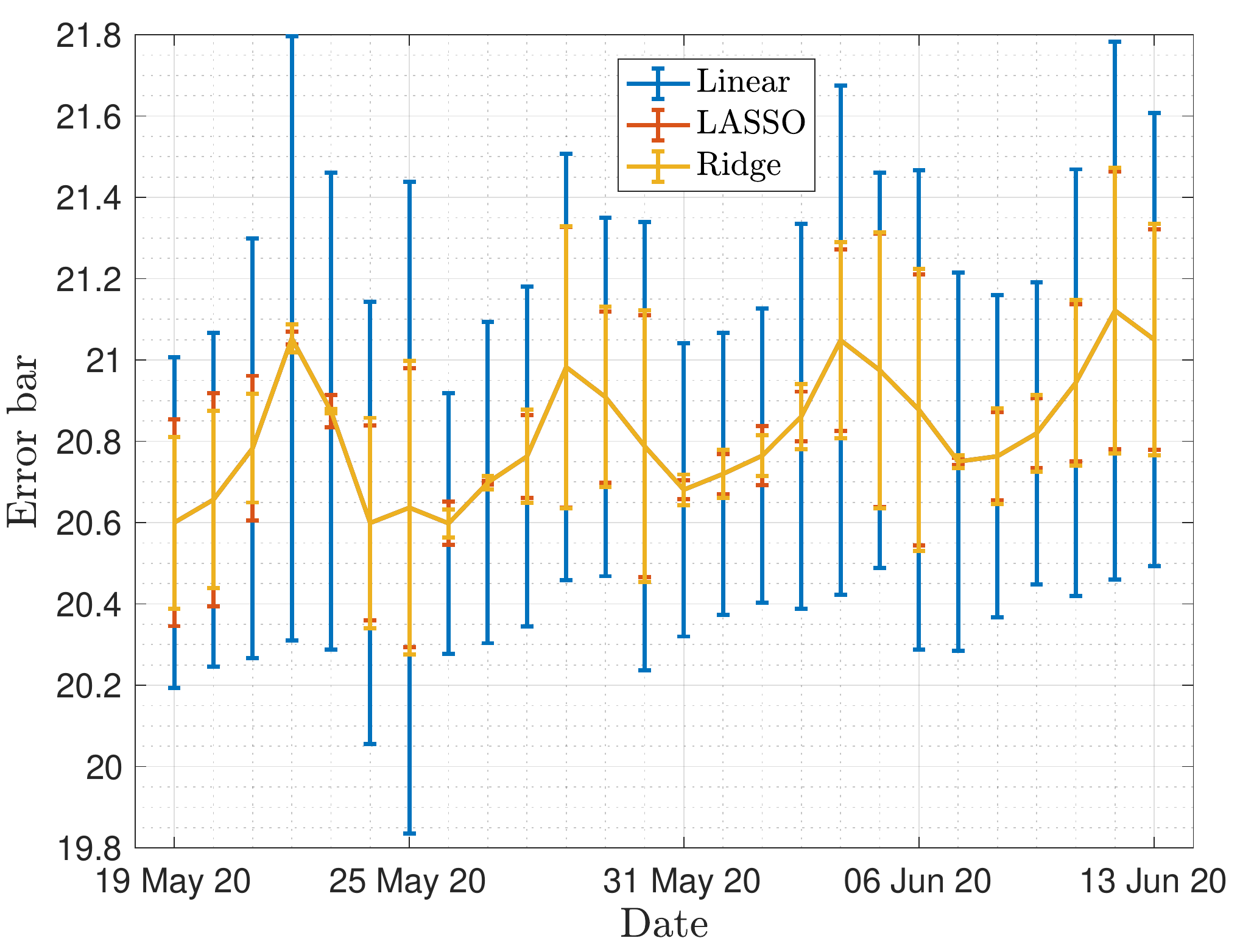}}
\caption{Error in prediction of daily VMT in the states of (a) Florida, (b) New Jersey, (c) New York and (d) Texas .}\label{rmse_sample_states}
\end{figure}

For the purpose of showing how the different modeling techniques perform at an individual state level, a comparative graphical representation of RMSEs in the prediction of daily VMT for the three regression techniques are presented separately for the states of Florida, New Jersey, New York, and Texas as examples in Figure \ref{rmse_sample_states}.  These figures clearly show that even at the individual state level, normal linear model provided the poorest performance in terms of large errors, while Ridge and LASSO models showed comparable prediction performance.  Among the three models, Ridge showed the most superior performance, followed by LASSO regression with a slightly higher prediction error.

\begin{figure}[htp!]
\centering
\subfigure[]{\includegraphics[scale=.39]{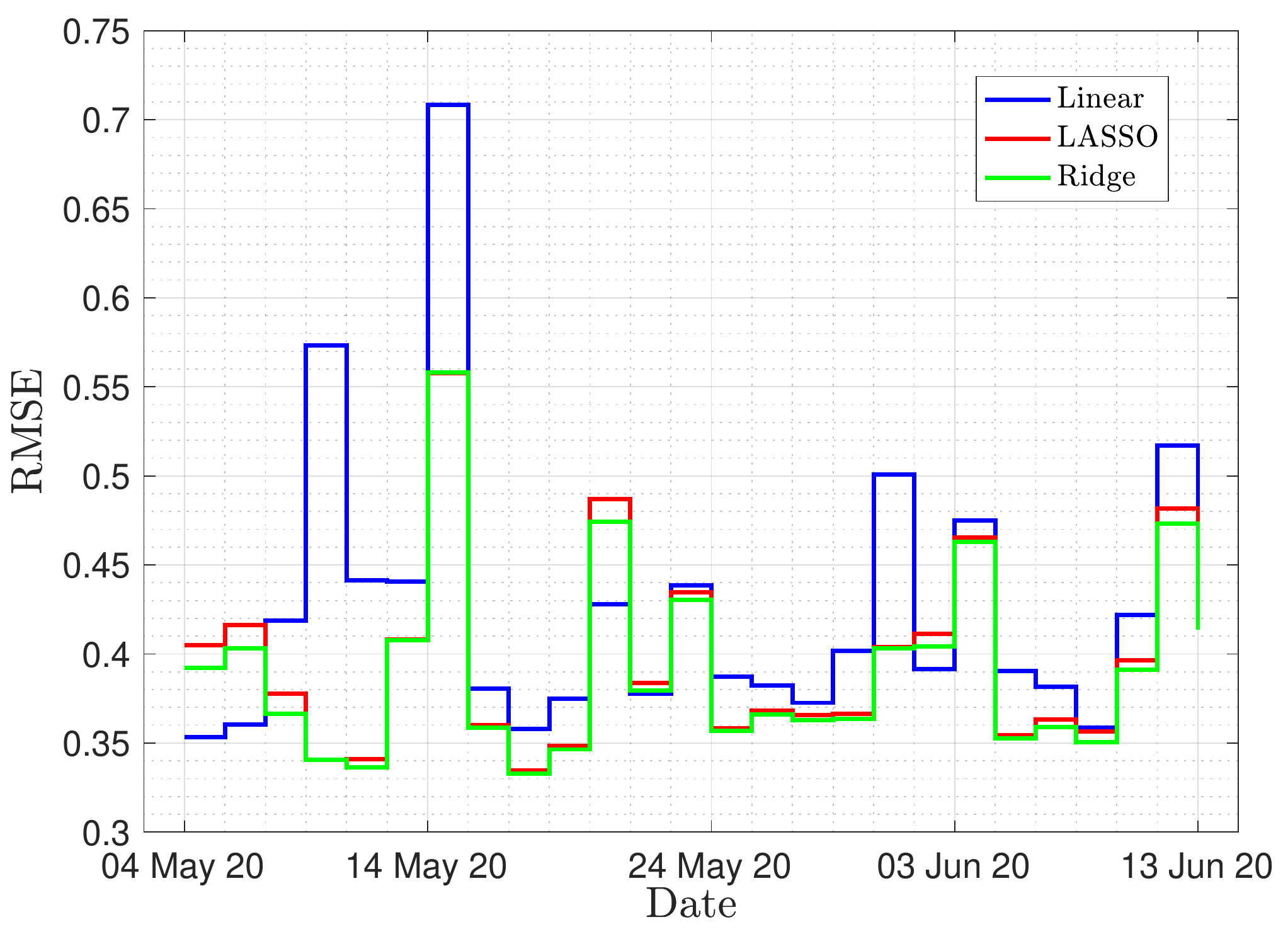}}
\subfigure[]{\includegraphics[scale=.39]{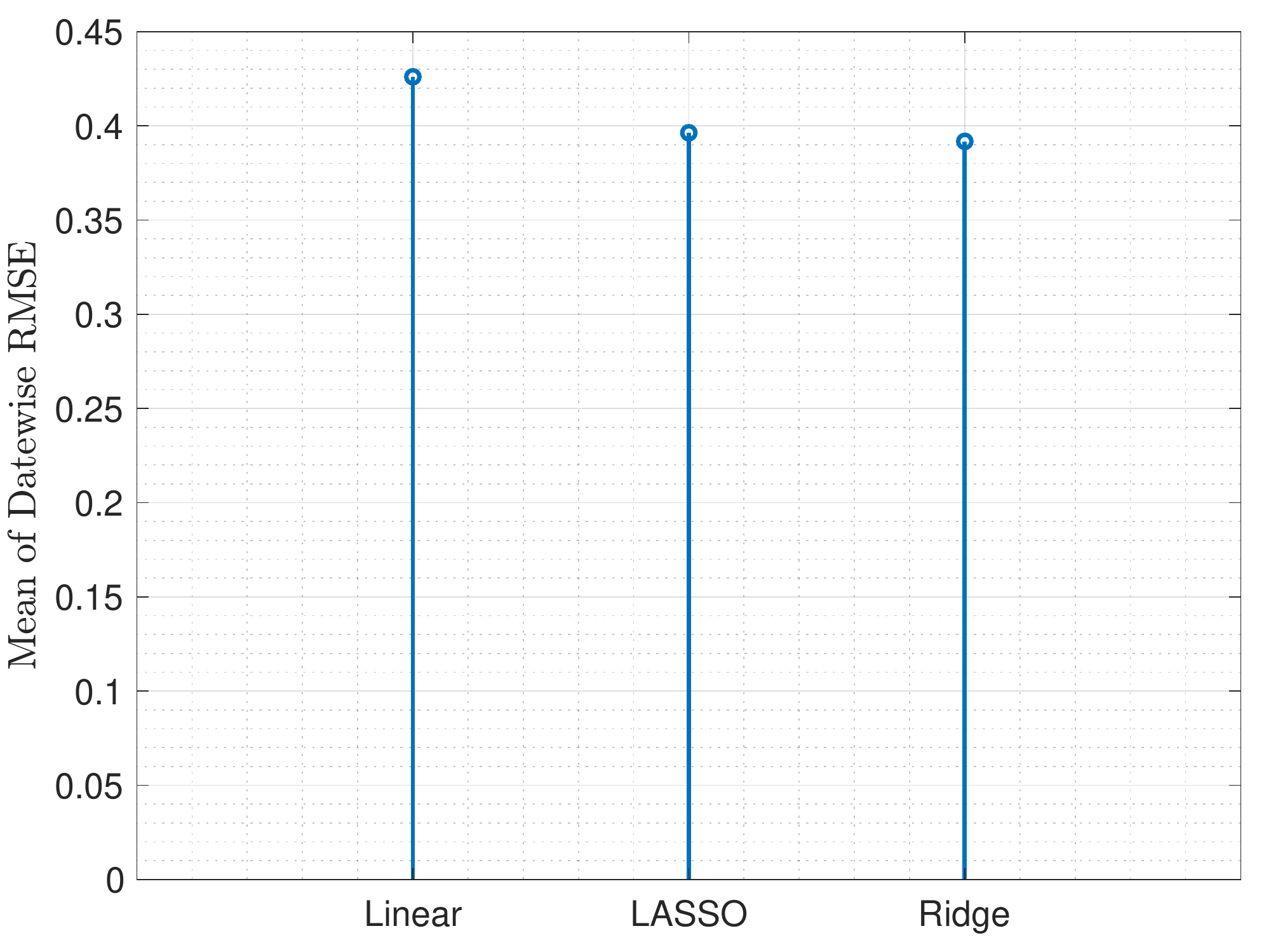}}
\caption{(a) Root mean square error (RMSE) in prediction of the different models (Eq. (\ref{rmse_prediction_datewise})) over the analysis period. (b) Average date-wise root mean square error (RMSE).}\label{rmse_datewise}
\end{figure}

Additionally, the date-wise RMSEs in the prediction, utilizing the three developed models over the analysis period are presented in Figure \ref{rmse_datewise}.  Let $y_{d_i}$ be the observed value of daily VMT on the $d^{th}$ day for the state $i$.  For the test data set, $d = 1,2,\cdots , 26$ and $i=1,2, \cdots , 49$.  Let $\hat{y}_{d_i}^p$ be the predicted daily VMT on the $d^{th}$ day for $i^{th}$ state when using the $p^{th}$ modeling technique.  Here $p$ is one of linear, LASSO or Ridge regression techniques.  Therefore, the error in the prediction using the $p^{th}$ model on the $d^{th}$ day for $i^{th}$ state is,
\vspace{-5mm}

\begin{eqnarray}
\epsilon_{d_i}^p = y_{d_i} - \hat{y}_{d_i}^p.
\end{eqnarray}

The RMSE in the prediction of the $p^{th}$ model for the $d^{th}$ day over all the states is, 

\vspace{-2mm}
\begin{eqnarray}\label{rmse_prediction_datewise}
r_{d}^p = \sqrt{\frac{\sum_{i=1}^{n_s}(\epsilon_{d_i}^p)^2}{n_s}}=\sqrt{\frac{\sum_{i=1}^{n_s}(y_{d_i} - \hat{y}_{d_i}^p)^2}{n_s}},
\end{eqnarray}
where $n_s = 49$ is the number of the states and D.C.

In Figure \ref{rmse_datewise}(a), $r_d^p$ is plotted for all modeling techniques over the period ranging between May 19, 2020 through June 13, 2020 (test period).  From Figure \ref{rmse_datewise}(a), it can once again be seen that the Ridge regression performed the best for most of the test period, as its RMSE was usually lower compared to the other methods.  The average of $r_{d}^p$ across the test period, which can be expressed as,

\begin{eqnarray}
r^p = \frac{\sum_{d=1}^{n_d}r_d^p}{n_d}=\left(\sum_{d=1}^{n_d}\sqrt{\frac{\sum_{i=1}^{n_s}(y_{d_i} - \hat{y}_{d_i}^p)^2}{n_s}}\right)\Bigg/ n_d,
\end{eqnarray}
where $n_d$ is the number of days (equals to 26 in this study), is plotted in Figure \ref{rmse_datewise}(b).  This plot also confirms that the Ridge regression model has the least error in prediction.

\section{Summary and Conclusions}\label{sec_conclusions}
Since the emergence and rapid growth of novel coronavirus (COVID-19) in December 2019, countries worldwide are taking extreme measures to prevent the spread of the virus.  The U.S. is greatly hit by the pandemic and currently (September 2020) has the highest number of confirmed COVID cases and deaths in the world.  Since the national emergency declared by the White House on March 13, 2020, most states in the U.S. implemented travel restrictions and social distancing protocols to combat the crisis, causing drastic reductions in mobility and travel demand at all levels.  However, the overall impact and the long-term implications of this crisis to mobility and travel still remain unknown at this point in time.  In order to understand these implications better, statistical models and analytical tools utilizing the increasingly available open-access data is the need of the hour.  To that end, this study developed an analytical framework that helped determine and analyze the most important factors impacting and predicting human mobility and travel in the U.S. during the pandemic by employing Granger causality and linear regularization algorithms, including the Ridge, and LASSO modeling techniques.

Data for this study were obtained for 48 states (excluding Alaska and Hawaii) and the District of Columbia from various databases created and maintained to analyze the impacts of this pandemic.  The data obtained and analyzed in this study included information on mobility and movement trends, travel restrictions and social distancing, and health and demographics of the population.  The compiled data set included daily time-series data starting from March 1, 2020 through June 13, 2020.

Evaluating a large-scale data set requires advanced analytical techniques to identify the most important factors in explaining the response variable.  Commensurate with analyzing such rich data, this study employed Granger causality and linear regularization techniques, including the Ridge, and LASSO models, along with ordinary least square regression.  The entire data set was split into two parts, where approximately 75 percent of the data was used for training the models, and the remaining 25 percent was used to test the prediction performance.  Determining the set of most important predictors impacting the daily VMT from the pool of several potential determinants was accomplished using the Granger causality.  The seventeen selected variables were further used to develop reduced order models, employing the linear, Ridge, and Lasso regression techniques on the training data.  Finally, the performance of the prediction was tested by feeding the test data into the models for all regression techniques.

The results of this study revealed that the coefficients of the predictors were comparable across all modeling techniques.  When factors including the number of new COVID cases, the social distancing index, percent of people working from home, statewide closure, and tests done per 100 people increased, the daily VMT decreased.  Conversely, the vehicle miles traveled per day increased with the increase in population, unemployment rate, person of out of county trips, socio-economic status, and the increase in percent of trips to transit stations, retail and recreational places, grocery and pharmacy, workplaces, and residences.  The population staying at home was rightly captured by regularization methods, namely LASSO and Ridge regressions, and shows a negative association with the daily VMT.

Furthermore, the developed models were used to predict the daily VMT for all the states for a period of 26 days (from May 19, 2020 through June 13, 2020).  Although all the developed models compare favorably, the Ridge regression model performed the best by having the least root mean square error (RMSE) in prediction among all models.  This result makes sense because the Ridge regression is robust in overcoming the issue of overfitting and thus generalizes better to the test data set, resulting in lesser prediction error.  Also, LASSO regularization technique performed superior to the ordinary least square regression. 

The study is only the starting point to help understand the associations between different factors and human mobility during the COVID-19 pandemic.  The authors of this study intend to expand this study to utilize county-based data to understand these associations from a more granular level.   Moreover, it would be insightful to include additional variables into the analysis as potential predictors.  From the modeling perspective, it is reasonable to argue that the available data is subjected to some uncertainties and future research should be carried out to explicitly take the uncertainties into account to derive at more precise models.  Furthermore, as the crisis is moving on to the greater peaks in terms of the number of confirmed cases and deaths over time in the U.S., subsequent analysis is warranted with data from the following months (post June 13, 2020).

\section{Acknowledgement}
None. This research did not receive any specific grant from funding agencies in the public, commercial, or not-for-profit sectors.

\section{CRediT Author Statement}
Subhrajit Sinha: Conceptualization, Methodology, Validation, Formal Analysis, Investigation, Data Curation, Original Draft Preparation, Review and Editing.  Meghna Chakraborty: Methodology, Validation, Formal Analysis, Investigation, Data Curation, Original Draft Preparation, Review and Editing.

\bibliographystyle{elsarticle-num-names}
\bibliography{sample}







\end{document}